\documentclass[final,3p,times]{elsarticle}

\usepackage[numbers]{natbib}
\biboptions{sort&compress}
\usepackage{latexsym}
\usepackage{amssymb}
\usepackage[mathscr]{eucal}
\usepackage{epsfig}
\usepackage{amsthm}
\newcommand{\be}{\begin{equation}}
\newcommand{\ee}{\end{equation}}
\newcommand{\bea}{\begin{eqnarray}}
\newcommand{\eea}{\end{eqnarray}}

\newtheorem{Theorem}{Theorem}

\newtheorem{Lemma}[Theorem]{Lemma}

\journal{Applied Numerical Mathematics}

\begin{document}

\begin{frontmatter}

\title{On the ambiguity of functions represented by divergent power series}

 \author[IC]{Irinel Caprini}
\ead{caprini@theory.nipne.ro}
 \address[IC]{National Institute of Physics and Nuclear Engineering, Bucharest POB MG-6, R-077125 Romania}
 \author[JF]{Jan Fischer\corref{cor1}}
 \ead{fischer@fzu.cz}
\cortext[cor1]{Corresponding author}
 \address[JF]{Institute of Physics, Academy of Sciences of the Czech Republic,
CZ-182 21  Prague 8, Czech Republic}
\author[IV]{Ivo Vrko\v{c}} 
\ead{vrkoc@math.cas.cz}
\address[IV]{Mathematical Institute, Academy of Sciences of the Czech
Republic,CZ-115 67  Prague 1,  Czech Republic}
\begin{abstract}
Assuming the asymptotic character of divergent perturbation series, we 
address the problem of ambiguity of a function determined by an asymptotic 
power expansion. We consider functions represented by an integral of the 
Laplace-Borel type, with a curvilinear integration contour. This paper is a 
continuation of results recently obtained by us in a previous work.
  Our new result contained in Lemma 
3 of the present paper represents a further extension of the class of contours 
of integration (and, by this, of the class of functions possessing a given 
asymptotic expansion), allowing the curves to intersect themselves or return 
back, closer to the origin. Estimates on the remainders are obtained for 
different types of contours. Methods of reducing the ambiguity by additional 
inputs are discussed using the particular case of the Adler function in QCD.
\end{abstract}


\end{frontmatter}

\section{Introduction}
\label{sec:intro}
In 1952, Freeman Dyson obtained the famous result \cite{Dyson} that 
perturbation series in QED are divergent. During the subsequent decades, 
similar results have been obtained \cite{Lautrup, Lipatov,Mueller, Parisi, 
tHooft} for most of the physically interesting field theories and models in 
quantum physics (for a review, see \cite{Fischer1,Fischer2} and references 
therein). This result was a surprise and set a challenge for a radical 
reformulation of perturbation theory. Dyson's suggestion to regard a divergent 
perturbation series as asymptotic has been universally adopted. Now the problem 
is not whether a perturbation series is convergent or divergent, but rather 
whether or not, and under what conditions, it uniquely determines the expanded 
function. A crucial task is to find effective additional inputs that would be 
able to reduce or, if possible, remove the ambiguity. If all expansion 
coefficients are known, the series may determine the sought function even if it 
is not convergent, and may not do so even if it is convergent. This depends on 
additional conditions.

How to deal with divergent series and how to sum them, under what conditions 
a power series is able to determine uniquely the expanded function and how to 
give a series a precise meaning are problems of paramount importance in quantum 
theory. Power expansions are badly needed in physics but, to ensure that they 
have clear mathematical meaning, additional conditions are necessary, which are 
often difficult to fulfill. 

We discuss here the ambiguities of perturbation theory stemming from the 
(assumed) asymptotic character of the series. We recall in section 
\ref{sec:watlem} the  Lemma of Watson (calling it Lemma 1). Then, in section 
\ref{sec:modwat}, we briefly recall our Lemma 2, which we obtained and proved 
in ref. \cite{Caprini1} for curvilinear contours of integration. Section 
\ref{sec:turn}  is a new result: we obtain and prove Lemma 3, which deals with 
certain specific forms of curvilinear integration contours.
 
 We shall use the following definition of asymptotic series. Given a point set ${\cal S}$  having the origin
 as a point of accumulation, the power series $\sum_{n=0}^{\infty}F_{n}z^{n}$ is said to be asymptotic to 
 the function $F(z)$ as $z \to 0$ on ${\cal S}$, 
if the set of functions $R_{N}(z)$,
\begin{equation}
R_{N}(z) = F(z) - \sum_{n=0}^{N}F_{n}z^{n} ,
\label{rema}
\end{equation}
satisfies the condition
\begin{equation}
R_{N}(z) = o(z^{N}) 
\label{ordo}
\end{equation} 
for all $N=0,1,2,...$, $z \rightarrow 0$ and $z \in {\cal S}$. The standard notation for an asymptotic series is:
\begin{equation}
F(z) \,\, \sim  \,\, \sum_{n=0}^{\infty} F_{n} z^{n},\quad\quad\quad  
z \in {\cal S}, \quad  z \rightarrow 0.
\label{Nevan1}
\end{equation}

The function $F(z)$ may be singular at $z=0$. The coefficients $F_{n}$ 
can be defined by
\begin{equation}
F_{n} = \lim_{z \to 0, z \in {\cal S}}   \frac{1}{z^{n}}  \left[F(z) - 
\sum_{k=0}^{n-1}F_{k}z^{k} \right],\quad  n = 0, 1, 2, ...
\label{coeff2}
\end{equation}
where $\sum_{k=0}^{n-1}F_{k}z^{k}=0$ for $n=0$ by definition. The prescription
(\ref{coeff2}) makes sense whenever the asymptotic expansion  exists: one can 
define $F_n$ without using the $n$-th derivative of $F(z)$, 
$z \in {\cal S}$, which may not exist. 

Let us recall that if the power series $\sum_{n=0}^{\infty}F_{n}z^{n}$ is 
convergent in a neighbourfood ${\cal L}$ of the origin and if $F(z)$ is
holomorphic and {\it equal} to $\sum_{n=0}^{\infty}F_{n}z^{n}$ in ${\cal L}$, 
then $F(z)$ is uniquely determined in ${\cal L}$. No additional input is needed,
in contrast with the case that the series is asymptotic. Asymptoticity can 
be checked only if one knows both the expansion coefficients and the expanded 
function $F(z)$.

 The ambiguity of a function given by an asymptotic series is illustrated  
 by Watson lemma.

\section{The lemma of Watson}	\label{sec:watlem}
Consider the following integral 
\begin{equation}
\Phi^{(\alpha,\beta)}_{0,c}(\lambda) = \int_{0}^{c} e^{-\lambda
x^{\alpha}}\,x^{\beta -1}f(x) {\rm d}x , 
\label{Laplace}
\end{equation}
where $0<c<\infty$ and  $\alpha > 0, \,\beta > 0 $. Let $f(x) \in 
C^{\infty}{[0,c]}$ and $f^{(k)}(0)$  defined as $\lim_{x \to 0+} f^{(k)}(x)$.
Let $\varepsilon$ be any number from the interval $(0, \, \pi/2)$. 

 \medskip
\begin{Lemma}\label{lemma1} (G.N. Watson) If the above conditions are 
fulfilled, the asymptotic expansion
\begin{equation}
\Phi^{(\alpha,\beta)}_{0,c}(\lambda) \sim \frac{1}{\alpha}\sum_{k=0}^{\infty} 
\lambda^{-\frac{k+\beta}{\alpha}}\, \Gamma
\bigg(\frac{k+\beta}{\alpha}\bigg)\frac{f^{(k)}(0)}{k!} 
\label{Watson}
\end{equation}
 {holds \em for $\lambda \rightarrow \infty, 
\lambda \in S_{\varepsilon}$, where  $S_{\varepsilon}$ is the sectorial region}
\begin{equation}
|\arg \lambda| \leq \frac{\pi}{2} - \varepsilon < \frac{\pi}{2}.
\label{uhel}
\end{equation}
 The expansion (\ref{Watson}) can be differentiated with 
respect to $\lambda$ any number of times.
\end{Lemma} 

 For the proof see for instance \cite{Jeff, Dingle, Fedo}.
 
\medskip It is worth mentioning that the region (\ref{uhel}) is independent 
of $\alpha, \,\beta$ and $c$, and the expansion coefficients 
in (\ref{Watson}) do not  depend on $c$. Further,  the factor 
$\Gamma \bigg(\frac{k+\beta}{\alpha}\bigg)$ in (\ref{Watson}) makes the 
expansion coefficients grow faster with $k$ than those of the Taylor series
for $f(x)$. For $\alpha=\beta=1$, $\Gamma \bigg(\frac{k+\beta}{\alpha}\bigg)$ 
in (\ref{Watson}) cancels with the factorial $k!$ in the denominator. 


The integral (\ref{Laplace}) \label{} reveals the large  ambiguity  of the 
resummation procedures having the same asymptotic expansion. No particular 
value $c$ of the upper limit of integration can be a priori preferred.

\medskip Below we shall recall our Lemma \ref{lemma2} (stated  and proved in 
ref. \cite{Caprini1}) showing a set of plausible conditions under which the 
integration contour in the Laplace-Borel integral can be bent. 

\section{Bending the integration contour}	\label{sec:modwat}
Let  $G(r)$  be a continuous complex function $G(r) = r \exp (ig(r))$, where 
$g(r)$ is a real-valued function given on $0\le r < c$, with   
$0<c \leq \infty$. Assume that the derivative $G'(r)$ is continuous on 
the interval $0\le r < c$ and a constant $r_0>0$ exists such that 
\begin{equation}
|G'(r)| \le K_1 r^{\gamma_1}, \quad\quad r_0\le r < c,
\label{CK1}
\end{equation}
for a nonnegative $K_1$ and a real $\gamma_1$.

Let the constants $\alpha>0$ and $\beta>0$ be given and assume that the 
quantities
\begin{equation}
A = \inf_{r_0\leq r < c}\alpha  g(r),\quad\quad \quad B = 
\sup_{r_0 \leq r < c}\alpha g(r)
\label{infsup} 
\end{equation}
satisfy  the inequality
\begin{equation}
B-A < \pi-2\varepsilon,
\label{eps} 
\end{equation}
 where $\varepsilon >0$.

Let the function $f(u)$ be defined along the curve $u=G(r)$ and on the disc 
$|u|<\rho$, where $\rho>r_0$. Assume $f(u)$ to be holomorphic on the disc 
and measurable on the curve. Assume that  
\begin{equation}
|f(G(r))|\leq K_2 r^{\gamma_2},\quad\quad r_0\le r < c,
\label{CK2}\end{equation}
hold for a nonnegative $K_2$ and a real $\gamma_2$.

Define the function $\Phi_{b,c}^{(\alpha,\beta,G)}(\lambda)$ for $0 \leq b< c$
by
\begin{equation}
\Phi_{b,c}^{(\alpha,\beta,G)}(\lambda)= \int_{r=b}^{c} e^{-\lambda (G(r))^{\alpha}} 
(G(r))^{\beta-1} f(G(r)) dG(r).
\label{bc}
\end{equation}
This integral exists since we assume $f(u)$ measurable along
the curve $u=G(r)$ and bounded by (\ref{CK2}).

\medskip
\begin{Lemma}\label{lemma2} If the above assumptions are fulfilled, then 
the asymptotic expansion
\begin{equation}
\Phi_{0,c}^{(\alpha,\beta,G)}(\lambda) \sim \frac{1}{\alpha} \sum_{k=0}^\infty
\lambda^{-\frac{k+\beta}{\alpha}}\, \Gamma \bigg(\frac{k+\beta}{\alpha} \bigg)
\frac{f^{(k)}(0)}{k!}  
\label{V}
\end{equation}
  holds for $\lambda \rightarrow \infty, \lambda \in \cal T_{\varepsilon}$, 
  where
\begin{equation}
{\cal T}_\varepsilon = \{\lambda: \lambda=|\lambda| \exp({\rm i} \varphi), \, 
\, \, - \frac{\pi}{2}- A + \varepsilon <\varphi< \frac{\pi}{2} - B- 
\varepsilon  \}.
\label{calT}
\end{equation}
\end{Lemma}

We refer the reader to our recent paper \cite{Caprini1}, where Lemma 2 is 
proved. The aim 
of the present paper is to show that a further generalization is possible. 
We shall show in section \ref{sec:turn} that  Lemma 2 in Ref. \cite{Caprini1} 
can be improved to apply to a wider class of curvilinear contours, including 
those that were mentioned in Remark 9 of ref. \cite{Caprini1}. According to 
that Remark, the parametrization $G(r)=r \exp{(ig(r))}$ does not include 
contours that cross a circle centred at $r=0$ either touching or doubly 
intersecting it, so that the derivative $G'(r)$ does not exist or is not 
bounded. In particular, this parametrization does not include the contours 
\begin{itemize}
\item that, starting from the origin and reaching a value $r_1$ of $r$, return 
back to a certain value $r_2<r_1$, which is closer to the origin, and
\item whose one or several parts coincide with a part of a circle centred at 
the origin, and
\item that have, at some point, their tangent perpendicular to the radius
vector. 
\end{itemize}

In the following section 4, we shall discuss a result that 
generalizes some features of Lemma 2 and, among others, cover also the two items mentioned above. 
For simplicity,  we  limit ourselves to $\alpha=\beta=1$ and $c$ finite. We shall 
call this new result Lemma \ref{lemma3}.
\section{Allowing the contour to cirsumscribe a circle or get closer towards 
origin}\label{sec:turn}
Let a complex function $G(s)$ be given. It is a function of a real parameter
$s$ on an interval $[0,c],c < \infty$. Assume that $G(s)$ has continuous
derivatives on $[0,c], G(0)=0, G(s) \neq 0$ for any $s>0$. Let the function 
$f(u)$ be defined along the curve $u = G(s)$ and on the disc ${\cal K}$ defined
by $|u|<\rho $, where $\rho>0 $. Assume $f(u)$ to be holomorphic 
on ${\cal K}$  and measurable and bounded on the curve. This implies
\begin{equation}
|f(G(s))| \leq K_2  \quad \mbox{for} \quad s \in [0,c].  
\label{gen2}
\end{equation}
 We choose a number $s_1$ such that $0<s_1<c $ and $G(s)$ lies in ${\cal K}$ 
for $s \in [0,s_1]$. Define
\begin{equation}
A=\inf_{s_1 \leq s \leq c} \arg(G(s)), \quad B=\sup_{s_1 \leq
s \leq c} \arg(G(s))
\label{gen4}
\end{equation}
and assume that
\begin{equation}
B-A< \pi-2 \varepsilon 
\label{gen5}
\end{equation}
where $\varepsilon >0$. Denote
\begin{equation}
{\cal U}_{\varepsilon} = \bigl\{\lambda : -\frac{\pi}{2}-A+ \varepsilon
<\arg[\lambda]< \frac{\pi}{2}-B- \varepsilon \bigr\} .  
\label{gen6}
\end{equation}

 Define the function
\begin{equation}
\Phi_{a,b}^{(G)}(\lambda)= \int_{s=a}^{s=b} f(G(s))
e^{-\lambda (G(s))} dG(s), 
\label{gen7}
\end{equation}
for $0 \leq a<b \leq c$, where the suppression of the labels $\alpha$ and
$\beta$ indicates that we have chosen $\alpha=\beta=1$. Note that we introduce 
here $s$, a new real variable, which parametrizes the length of the integration
contour and, unlike $r$, does not mean the distance from the origin. 

\begin{Lemma}\label{lemma3}  If the above assumptions are fulfilled, then 
the asymptotic expansion
\begin{equation}
\Phi_{0,c}^{(G)}(\lambda) \sim \sum_{k=0}^\infty \lambda^{-(k+1)} f^{(k)}(0)
\label{gen8}
\end{equation}
 holds for $\lambda \rightarrow \infty, \lambda \in
{\cal U}_ {\varepsilon} $.
\end{Lemma}

\noindent
{\bf Remarks:}   

\medskip
1. The cone ${\cal U}_{\varepsilon}$ (\ref{gen6}) is maximal. It is proved in 
\cite{Caprini1} that outside ${\cal T}_\varepsilon$ Lemma \ref{lemma2} might 
not be  fulfilled. The same argument can be applied in the case of Lemma 
\ref{lemma3}. 

2. If a curve is rectifiable and of finite length, then the value of $s$ for a
point of the curve can be defined as a function of the length of the curve 
from the origin to that point.

\medskip
\noindent
{\bf Proof:} For a given $N$, $f(u)$ can be expressed inside the 
circle of radius $\rho' < \rho$ in the form
\begin{equation}
f(u) = \sum_{k=0}^N\frac{f^{(k)}(0)}{k!} u^k +r_N(u), \quad
|r_N(u)| \leq C_N
|u|^{N+1}. 
\label{gen10}
\end{equation} 
Since the interval [0,c] is compact and $G'(s)$ is continuous there exists a 
constant $K_1$ such that
\begin{equation}
|G'(s)| \leq K_1 \quad \mbox{for} \quad s \in [0,c]. 
\label{gen1}
\end{equation}
Further, there exists a positive number $\eta$ such that
\begin{equation}
|G(s)| > \eta \quad \mbox{for} \quad s \in [s_1,c]  
\label{gen3}
\end{equation}
(note that $G(s) \ne 0$ in $[s_1,c]$ because $s_{1} > 0$).

\medskip Let us define 
\begin{equation}
\tilde G(s) =\frac{s}{s_1}\, G(s_1) \quad \mbox{for} \quad 0 \leq s
\leq s_1, \quad \tilde G(s) = G(s) \quad \mbox{for} \quad
 s > s_1. 
 \label{gen11}
\end{equation}    
Since the curves $\tilde G(s),G(s)$ lie in ${\cal K}$ for $0 \leq s \leq
s_1$ (note that $f(u)$ is holomorphic in ${\cal K}$) and $\tilde G(0)=G(0)$, 
$\tilde G(s_1)= G(s_1)$, the integrals of the function $ f(u) e^{-\lambda u}$ 
along these curves on the interval $[0, s_1]$ equal each other. We have 
\begin{equation}
\Phi_{0,c}^{(G)}(\lambda)=\Phi_{0,s_1}^{(\tilde G)}(\lambda)+
\Phi_{s_1,c}^{(G)}(\lambda). 
\label{gen12}
\end{equation}
Let us define 
\begin{equation}
I_{a,b}^k(\lambda) =
\int_{a}^b \left(\frac{s \ G(s_1)}{s_1}\right)^k \exp{\{-\lambda(s/s_1) 
G(s_1)\}} \frac{1}{s_1} G(s_1)\, ds, 
\label{gen13}
\end{equation}
where the integrals run along the ray  $(s/s_1) G(s_1)$. We obtain, using
(\ref{gen11}), 
\begin{equation}
\Phi_{0,s_1}^{(\tilde G)}(\lambda)=
\sum_{k=0}^N I_{0,s_1}^k(\lambda)f^{(k)}(0)/k!+
\int_{s=0}^{s=s_1} r_N(\tilde G(s)) \exp{\{-\lambda \tilde G(s)\}}d\tilde G(s).
\label{gen14}
\end{equation} 
Let us first calculate the terms $I_{0,s_1}^k(\lambda)$.
We have 
\begin{equation}
I_{0,s_1}^k(\lambda)=I_{0,\infty}^k(\lambda)-
I_{s_1,\infty}^k(\lambda). 
\label{gen15}
\end{equation}
To calculate the first term $I_{0,\infty}^k(\lambda)$, we shall use 
the well-known formula:
\begin{equation} 
\int_0^ \infty x^{\delta-1} \exp{\{-\mu x\}}
dx=\frac{1}{\mu^\delta}\Gamma(\delta)  
\label{gen16}
\end{equation}
for Re $\delta >0$, Re $\mu>0 $. 
If we take $\delta=k+1, \mu= (\lambda/s_1)G(s_1)$, we obtain 
\begin{equation}
I_{0,\infty}^k=\frac{1}{\lambda^{k+1}} \Gamma(k+1).  
\label{gen17}
\end{equation}
We shall show that, for the last term in (\ref{gen12}), the following 
inequality holds:
\begin{equation}
|\Phi_{s_1,c}^{(G)}(\lambda)| \leq
K_1 K_2 c \exp{\{-|\lambda|\eta \sin \varepsilon\}},  
\label{gen18}
\end{equation}
which is an exponential estimate. 

Having chosen the cone ${\cal U_\varepsilon}$ (\ref{gen6}) and using 
(\ref{gen3}), we obtain for $\lambda \in {\cal U}_\varepsilon$ the inequality 
\begin{equation}
{\rm Re}[\lambda G(s)]
\geq |\lambda| |G(s)| \cos[\arg \lambda+\arg(G(s)]
\geq |\lambda| \eta  \sin \varepsilon. 
\label{gen19}
\end{equation}
Hence 
\begin{equation}
|e^{-\lambda G(s)}|=e^{-{\rm Re}[\lambda G(s)]} 
\leq e^{-|\lambda| \eta \sin \varepsilon}. 
\label{gen20}
\end{equation}
The inequalities hold for s from the interval $[s_1,c]$. Further, 
\begin{equation}
|\Phi^{(G)}_{s_1,c} (\lambda)| \leq K_1 K_2 \int_{s_1} ^c
e^{-|\lambda| \eta \sin \varepsilon}  ds. 
\label{gen21}
\end{equation}
This proves that the estimate  (\ref{gen18}) holds.

 Now we shall deal with  $I_{s_1,\infty}^k(\lambda)$ (see (\ref{gen15})). 
 We have 
\begin{equation}
 |I_{s_1,\infty}^k(\lambda)| \leq \left(\frac{|G(s_1)|}{s_1}\right)^{k+1}
\int_{s_1}^\infty s^k \exp{\{-|\lambda| s/s_1 |G(s_1)|
\sin \varepsilon\}} ds.
\label{gen22}
\end{equation} 
The right hand side can be rewritten
\begin{equation}
\frac{1}{|\lambda|^{k+1}(\sin \varepsilon)^{k+1}}\int_{|\lambda
G(s_{1})|\sin \varepsilon}^{\infty} y^{k} e^{-y}dy .
\label{23}
\end{equation}

Certainly
\begin{equation}
|I_{s_1,\infty}^k(\lambda)| \leq  
\frac{K_{k,\delta}}{|\lambda|^{k+1}(\sin \varepsilon)^{k+1}(1-\delta)}
\exp{\{-(1-\delta)|\lambda G(s_1)|\sin \varepsilon\}} 
\label{gen23}
\end{equation}
for $|\lambda|>1$, where $0<\delta<1$ and $y^k < K_{k,\delta} 
{\rm e}^{\delta y}$ for $y > |G(s_1)|\sin \varepsilon$,  
which is again an exponential estimate. 

 The integral containing the remainder $r_N(z)$ (see (\ref{gen14}))
can be estimated in a similar way using the inequality  
\begin{equation}
\int_0^{s_1} (s \ |G(s_1)|/s_1)^{N+1} \exp{\{-|\lambda |s/s_1
|G(s_1)| \sin \varepsilon\}}1/s_1 |G(s_1)| {\rm d}s \leq 
\frac{\Gamma(N+2)}{|\lambda|^{N+2}(\sin \varepsilon)^{N+2}},  
\label{gen25}
\end{equation}
which implies that the second term on the right hand side of (\ref{gen14})
satisfies the inequality
\begin{equation}
\left\vert \int_{0}^{s_1} r_N(\tilde G(s)) \exp{\{-\lambda \tilde G(s)\}}d\tilde G(s)\right\vert 
\leq  C_N \frac{\Gamma(N+2)}{|\lambda|^{N+2}(\sin \varepsilon)^{N+2}}.
\label{estimate} 
\end{equation} 
This is a polynomial estimate of a lower degree than $I_{0,\infty}^k(\lambda)$.

\section{Discussion}	\label{sec:dscssn}
Lemma 3 and its proof cover up a set of integration contours that are not
embraced in Lemma 2. In both cases, 
the contour of integration starts at the origin, $u=0$. On the other hand, 
while the conditions of Lemma 2 admit only integration contours with increasing 
distance from the origin, the conditions of Lemma 3 permit a portion of the 
contour to get closer to the origin, or to have the form of an arc 
centred at the origin.  Also, in Lemma 3, the 
integration contour may both perform spirals  and intersect itself any 
number of times, with the reservation that the contour must not circle round 
the origin. It is a fundamental  feature of both Lemma 2 and Lemma 3 that the 
integration contour of the Borel-Laplace integral must not leave the sectorial region
${\cal T_{\varepsilon}}$ and ${\cal U_{\varepsilon}}$ respectively.  

\begin{figure}\begin{center}\vspace{0.2cm}
\includegraphics[width=7.5cm]{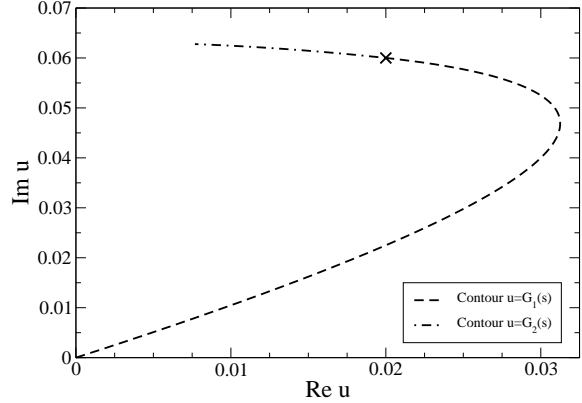}
\caption{\label{fig:Curves} An integration contour allowed by Lemma 3, but which
 does not satisfy the conditions of Lemma 2. The cross marks the transition between the two curves
discussed in the text.} \end{center}\end{figure}
  To illustrate the above remarks we consider a simple example: let the curve $u=G_1(s)$ 
in the $u$-plane be 
defined parametrically by 
\bea\label{G1s}
&&G_1(s)=t(s) + i \, v(s),\quad\quad  s\in [0, 1], \nonumber\\ 
&&t(s)=a_1 s + a_2 s^2, \quad\quad v(s)=b_1 s + b_2 s^2, 
\eea
where $a_1, a_2, b_1, b_2$ are real parameters.
It is easy to see that this curve satisfies the conditions of Lemma 3. On the 
other hand, it cannot be written always as $u=r\,\exp (i g(r))$, where $g(r)$ 
is a real function with a continuous first derivative, as requires Lemma 2. 
Indeed, let us make the change of variable
\begin{equation}\label{rs}
r\equiv r(s)=\sqrt{t(s)^2+v(s)^2}.
 \end{equation} 
Then
\begin{equation}\label{gr}
g(r)=\arctan[v(s(r))/t(s(r))],
 \end{equation}
where $s(r)$ is the inverse of (\ref{rs}). The derivative of (\ref{gr}) with 
respect to $r$ can be written as
\begin{equation}\label{gr1}
g'(r)=[v(s(r)) t'(s(r))-v'(s(r)) t(s(r))]\,\frac{s'(r)}{r^2} \,  
 \end{equation}
 where $s'(r)=1/r'(s)$. One can easily check that, for the choice $a_1 >0$,
 $0< b_{1} < 2 a_{1}$, and  
\begin{equation}\label{a2b2}
 a_2=- \frac{3 a_1 + b_1}{5}, \quad\quad b_2=\frac{a_1 -3 b_1}{5}.
\end{equation} 
 one has $r'(s)>0$ for $0<s<1$ and $r'(1)=0$. Then, (\ref{gr}) is
 justified because $G_{1}(s)$ lies in the first quadrant. 
 
 It follows that $s'(r) \to \infty$ for $r \to\ r(1)$  and, since the first 
 factor in (\ref{gr1}) does not vanish at $r=r(1)$, $g'(1)$ is not bounded in
 the neighbourhood of $r=r(1)$, 
 {\em i.e.} $g(r)$ does not fulfill the conditions of Lemma 2. (There are curves
 that possess infinitely many such points.) 

 The curve $G_{1}(s)$ can be further continued by $u=G_2(s)$ in such a way that 
 the conditions of Lemma 3 are satisfied: 
\bea\label{G2s}
&&G_2(s)=t(s) + i \, v(s),\quad\quad  s\in [1, 1.2], \nonumber\\ 
&&t(s)=(a_1+a_2) \cos(s - 1) -(b_1 + b_2) \sin(s - 1), \quad v(s)= (a_1+a_2) \sin(s - 1)+(b_1 + b_2) \cos(s - 1).
\eea
For any values of $a_i$ and $b_i$ this curve is an arc of a circle centered at 0, therefore the derivative of $|G_2(s)|$ with respect to $s$ is zero, while for the contours allowed in Lemma 2 the derivative should be equal to 1.   In Fig. 1 we represent the union of the two curves discussed above, for the choice $a_1=b_1=0.1$ and $a_2$, $b_2$ defined in (\ref{a2b2}).

 \section{Reducing the ambiguity by additional inputs} \label{sec:remqcd}

To discuss some physical applications we take the Adler 
function in massless QCD as an example. The  Adler function ${\cal D}(s)$ 
(see \cite{Adler}) is assumed to be real analytic in the complex $s$-plane cut 
along the timelike axis. The renormalization-group improved expansion,
\be\label{Dpert}
{\cal D}(s) = D_1 \,\alpha_s(s)/\pi +  D_2 \,(\alpha_s(s)/\pi)^2 + 
D_3 \,(\alpha_s(s)/\pi)^3  + \ldots \,,  
\ee 
has an additional unphysical singularity due to the Landau pole in the running 
coupling $\alpha_s(s)$. According to present knowledge, (\ref{Dpert}) is 
divergent, the $D_n$ growing as $n!$ at large $n$ 
\cite{Mueller1992, Beneke, Broad, BenekePR, Dyson}, see also \cite{Bender,
Guillou} and \cite{Fischer2} and references therein. 

 To discuss the implications of Lemma 2 and Lemma 3, 
we define the Borel transform $B(u)$  by  \cite{Neubert}
\be\label{B}
B(u)= \sum\limits_{n\ge 0} b_n \,u^n,\quad\quad \quad 
b_n=\frac{D_{n+1}}{\beta_0^n\,n!}\,,
\ee
where $\beta_0$ is the first coefficient of the $\beta$ function governing
 the  renormalization group equation satisfied by the coupling.
It is usually assumed that the series (\ref{B}) is convergent on a disc of
nonvanishing radius (this result was rigorously proved by David et al. 
\cite{David} for the scalar $\varphi^4$ theory).
This is exactly what is required in Lemmas 2 and 3 for the Borel transform.

If we adopt the assumption that the series (\ref{Dpert}) is asymptotic, both 
Lemma 2 and Lemma 3 imply  a large freedom in recovering the true function 
from its perturbative coefficients. Indeed, taking for simplicity 
$\alpha=\beta=1$ in (\ref{bc}), we infer that all the  functions 
${\cal D}^G_{0,c}(s)$ of the form
\be\label{BDG}
{\cal D}^G_{0,c}(s)=\frac{1}{\beta_0} \int_{r=0}^c e^{-\frac{G(r)}{\beta_0\,
a(s)}}\, B(G(r))\, {\rm d}G(r) \, ,
\ee  
where $a(s)=\alpha_s(s)/\pi$,  admit the asymptotic expansion of the type
(\ref{Dpert}), provided
that the assumptions of Lemma 3 are fulfilled. This reveals the large ambiguity of the resummation
procedures having the same asymptotic expansion in perturbative QCD. No
particular function of the form ${\cal D}^G_{0,c}(s)$ can be a priori preferred
when looking for the true Adler function.

\subsection{Mathematical conditions for uniqueness}
For completeness, in  this section we shall review several criteria for removing
 the ambiguity of a function represented by an
 asymptotic expansion.  A powerful tool to reach uniqueness 
is provided by the Strong Asymptotic Conditions (SAC), which are conditions for 
Borel summability. The problem was investigated in 
many papers (see  \cite{Sokal}, \cite{Balser}, \cite{Sauzin} and references therein). 

 The SAC are commonly used in two versions, one being due to
G. Watson and another one due to F. Nevanlinna. Watson's version (\cite{Watson},
see also \cite{Sokal}) of the uniqueness criterion gives a sufficient condition 
for $F(z)$ to equal the Borel sum of its asymptotic Taylor series: 

{\it Watson's criterion}:\,\,\,\,{\em Assume $F(z)$ to be analytic in a sector 
$S_{\varepsilon,R}$, $|{\rm arg}\,z|<\pi/2+\varepsilon$, $|z|<R$, for some 
positive $\varepsilon$, and let $F(z)$ have the asymptotic expansion}
\begin{equation}
F(z) = \sum_{k=0}^{N-1}a_{k}z^{k} + R_{N}(z) ,  \quad  {\rm where}  \quad
|R_{N}(z)| \leq A \, \sigma^{N} N! \, |z|^{N} 
\label{Nevan2}
\end{equation}
{\em for $N=0, 1,2,...$ uniformly in $N$ and in $z$ in the sector. Then 
$F(z)$ is uniquely determined, being equal to}
\begin{equation}
h(z) = 
\frac{1}{z}\int_{0}^{\infty} {\rm e}^{-u/z}B(u){\rm d}u, \quad {\rm where}  
\quad B(u) \, = \, \sum_{n=0}^{\infty} \frac{a_{n}}{n!} u^{n}
 \label{borel}
 \end{equation}
{\em inside the circle}  Re $\, z^{-1}>1/R$. 

Note that $\varepsilon$ is positive. This condition, sometimes difficult to 
satisfy, can be modified to a refined and improved version, which is due to 
Nevanlinna (\cite{Nevan}, see also \cite{Sokal}). Nevanlinna's condition of 
Borel summability  is: 

{\it Nevanlinna's criterion}:\,\,\,\,{\em Let $F(z)$ be analytic in the circle 
$C_{R} = \{z: {\rm Re} \, z^{-1}>1/R \}$ and satisfy there the estimates 
(\ref{Nevan2}) for $N=0, 1,2,...$  uniformly in $N$ and in $z \in C_{R}$. }
Then $F(z)$ is uniquely determined and is equal to the function $h$ defined in 
(\ref{borel}).

Nevanlinna's criterion gives both a sufficient and a necessary summability
condition, see \cite{Nevan, Sokal}. Formally it is obtained from Watson's by 
replacing the sector $S_{\varepsilon,R}$ with the disc $C_{R'}$, where $R$ and 
$R'$ may be different.  

In other words, among all the functions $F(z)$ analytic in $C_{R}$ and 
possessing the asymptotic expansion (\ref{Nevan1}) there is only one function, 
$h(z)$, which satisfies the inequalities (\ref{Nevan2}) for all $N=0,1,2,...$.  
Thus, among all functions $F(z)$ satisfying the expansion (\ref{Nevan1}) there 
is one, $h(z)$, which is the best, in the sense that all the remainders 
$R_{N}(z)$, $N=0,1,2,..$ are the smallest possible in $C_{R}$.

Further progress was achieved by T. Carleman \cite{Carlem}. Carleman's 
theorem can be used to show that two analytic functions with the same asymptotic
expansion are identical. Some infinitely differentiable but non-analytic 
functions vanish identically in certain subsets of the complex plane. 
Carleman's theorem has the following form (see, e.g., \cite{ReedSim}):

{\it Carleman's theorem}:\,\, Let  $g$ be a function analytic inside the 
sector $S_{R} = \{z|, \, 0 \leq |z| \leq R, |\arg z| \leq \pi/2 \}$ and 
continuous on $S_{R}$. Assume that 
\begin{equation}
|g(z)| \leq b_{N} |z|^N
\label{carlem}
\end{equation}
for every $N$ and all $|z|$ inside the sector. If 
$\sum_{n=1}^{\infty} b_{n}^{-1/n} = \infty$, then $g$ is identical zero. 

The methods described above are effective 
ways to remove the infinite ambiguity by selecting the function $h(z)$, which 
is "the nearest" in the sense that the remainders $R_{N}(z)$ of all orders, 
$N=0, 1, 2,...$, see (\ref{Nevan2}) (or the function $g(z)$, see 
(\ref{carlem})), are the smallest possible in the respective region 
$S_{\varepsilon,R}$, $C_{R}$         
and $S_{R}$.	
Nearness is a natural criterion; on the other hand, it is not evident that 
nearness is always the best motivation from the physical point of view. It is 
worth discussing also other options. 
\subsection{Analyticity, its splendour and its dangerous points}
In problems of divergence and ambiguity, the knowledge of the singularities 
of ${\cal D}(s)$ and of $B(u)$  is of importance. Some information about 
 ${B(u)}$  follows from certain classes of Feynman diagrams, from 
renormalization theory and general nonperturbative arguments. Due to the 
singularities at $u$ positive, the series (\ref{Dpert}) is not Borel summable. 
Except renormalons and instanton-antiinstanton pairs (i.e., $u \geq 2$ and 
$u \leq -1$), no other singularities of $B(u)$  are known. It is usually 
assumed that, with the exception of the above-mentioned singularities along the positive and
the negative real semiaxes with a gap around the origin, $B(u)$ is holomorphic 
elsewhere. 

To treat the analyticity properties of $B(u)$, the method of optimal 
conformal mapping \cite{Ciulli} is very useful. Applications of this 
method to Lemma 2 and its merits are discussed in our previous paper 
\cite{Caprini1}; the applications to Lemma 3 go along the same line. We 
refer the reader to \cite{Caprini1} and references therein for details. 

On the other hand, a careless manipulation with the integration contour may 
have a destructive effect on the analyticity properties. In \cite{Howe,Brooks}, 
two different contours are chosen for the summation of some class 
of diagrams: one contour,  parallel and close to the positive semiaxis, is 
chosen for $a(s)>0$, while another one,  parallel and close to the negative 
semiaxis, is taken when $a(s)<0$. As proved in \cite{Caprini2}, analyticity is 
lost with this choice, the summation being only piecewise analytic in $s$. 

On the other hand, as shown in \cite{Caprini3, Caprini4}, the Borel summation 
with the Principal Value (PV) prescription of  the same class of diagrams  
admits an analytic continuation to the whole $s$-plane, being consistent 
with analyticity except for an unphysical cut along a segment of the 
space-like axis, related to the Landau pole.  In this sense, PV is 
an appropriate prescription.

\section{Concluding remarks} \label{sec:conrem} 
The main result of our work is Lemma 3  proved in section \ref{sec:turn}, which 
emphasizes the great ambiguity of functions represented by asymptotic power 
series. The result holds if the function  $f(u)$ (which corresponds to the 
Borel transform) is analytic on a  disc in the Borel plane and satisfies rather 
weak conditions outside the disc. Lemma 3 is an extension of Lemma 2 
formulated and proved by us in ref. \cite{Caprini1}, and briefly mentioned in 
section \ref{sec:modwat}  of  the present paper. Lemma 2 and Lemma 3 are 
valid for two different classes of integration contours in the integral 
representations of the functions with a prescribed asymptotic expansion. 

If applied to perturbation theory, Lemma 2 and Lemma 3 draw one's attention to 
the fact of a great ambiguity of the summation prescriptions that are allowed 
if the perturbation expansion is regarded as asymptotic. The contour of the 
integral representing the function of interest and the corresponding function 
$B(u)$ can be chosen very freely outside the convergence disc.

 Lemma 2 and Lemma 3 proved in ref. \cite{Caprini1} and, respectively, in this 
paper may also be useful in other branches of physics where the perturbation or
other series are divergent. 

\medskip
\noindent
{\bf Acknowledgments}

IC thanks Prof. Ji\v{r}\'i Ch\'yla and the Institute of Physics of the Czech 
Academy in Prague for hospitality. JF thanks Sorin Ciulli, G\'{e}rard Menessier 
and Jean Zinn-Justin for useful discussions. We are greatly indebted to the
 referees, whose comments have significantly contributed to a clear presentation 
of the results of this paper.  Work supported by CNCSIS in the 
Program Idei, Contract No. 464/2009, and by the Projects No. LA08015 of the 
Ministry of Education and AV0-Z10100502 of the Academy of Sciences of the 
Czech Republic. 
\medskip

\bibliographystyle{elsarticle-num} 
\bibliography{FischerANMrev}

\end{document}